\documentclass[12pt,draftcls,onecolumn,twoside]{IEEEtran}
\usepackage{listings}
\usepackage{amsmath}
\usepackage{graphicx}
\usepackage{amsfonts}
\usepackage{array}
\usepackage{amssymb}
\usepackage{helvet}
\usepackage{color}
\usepackage{rawfonts,amsfonts,psfrag,flushend,geometry}
\usepackage{bm}
\usepackage{float}
\usepackage{cite}
\usepackage{multirow}
\usepackage{slashbox}
\makeatletter
\def\hlinewd#1{%
  \noalign{\ifnum0=`}\fi\hrule \@height #1 \futurelet
   \reserved@a\@xhline}
\makeatother
\begin{document}
%
\title{Enhanced Multi-Parameter Cognitive Architecture for Future Wireless Communications}
%
%
%

\author{Kaiqing Zhang, Feifei Gao, and Qihui Wu
\thanks{K. Zhang and F. Gao are with Tsinghua National Laboratory for Information Science and Technology, Beijing, 10084, China, email:(zkq11@mails.tsinghua.edu.cn, feifeigao@ieee.org)}
\thanks{Q. Wu is with the College of Communications Engineering, PLA University of Science and Technology, Nanjing, 211101, China, email:(wqhqhw@163.com)}
}

\maketitle

\vspace{-20mm}
\begin{abstract}
The very original concept of cognitive radio (CR) raised by Mitola targets at all the environment parameters,  including those in physical layer, MAC layer, application layer as well as the information extracted from  reasoning. Hence the first CR is also referred to  as ``full cognitive radio''. However, due to its difficult implementation, FCC and Simon Haykin separately proposed a much more simplified definition, in which CR mainly detects one single parameter, i.e., spectrum occupancy, and is  also called as ``spectrum sensing cognitive radio''. With the rapid development of wireless communication,  the infrastructure of a wireless system becomes much more complicated while  the functionality at every node is desired  to be  as intelligent as possible, say the self-organized capability in the approaching 5G cellular networks. It is then interesting to re-look into Mitola's definition and think whether one could, besides obtaining the ``on/off'' status of the licensed user only,  achieve more parameters in a cognitive way.
In this article, we propose a new cognitive architecture targeting at multiple parameters in future cellular networks, which is a one step further towards the ``full cognition" compared to the most existing CR research. The new architecture is elaborated in  detailed stages, and three representative examples are provided based on the  recent research progress to illustrate the feasibility as well as the validity of the proposed architecture.
\end{abstract}

\vspace{-8mm}

%
\IEEEpeerreviewmaketitle

\section{Introduction}
%
%
%
%

The rapid development of wireless communications has engendered quick proliferation of media-rich mobile devices as well as significant enhancement of communication ability.
The actual mobile traffic in 2010 is five times greater than an official forecast made by the International Telecommunication Union (ITU) in 2005  and would continuously increase about 1000 times till the end of 2020 \cite{Concepts_in_5G}.

The system capacity could be enhanced by enlarging the network coverage (via relays, femtocell), increasing the space dimension (via massive MIMO), improving the bandwidth efficiency and energy efficiency (via cognitive radio technology,  green communications) \cite{Key_Technologies}.
On the other side, dynamic networks are adopted to speed up the service innovation in a more intelligent way. For example, the self-organization networks (SON) with artificial intelligence serves as a promising solution to address the challenges imposed by large-scale networks, e.g., the high cost of configuring and managing networks, the fluctuation nature of the available spectrum, and the diverse QoS requirements of various applications\cite{SON,SON_Swarm}.
In order to  realize these key technologies,  cross-layer parameters, e.g., spectrum occupancy, transmit power, modulation, constellation, channel coding, location, and cell edge, should be achieved and shared among nodes in the network. However, the current information exchanging scheme still relies on the cooperative feedback among nodes, which causes severe transmission overhead especially when the scale of the network and the amount of the data traffic are large. Hence, an intelligent way to cognitively obtain as many parameters as possible at every  node  could greatly enhance the network efficiency.

The terminology ``full cognitive radio" has already been proposed by Mitola \cite{MitolaFull}  in which the cognition targets at every possible parameter observable by a wireless node or network.
However, this ultimate goal of cognitive radio (CR)  only stays in the conceptual level, while the recent 10 years' studies in parameter achieving mainly focused on one single parameter, i.e., the occupancy of a particular spectrum. Various parameters depicting the network status, e.g., channel occupancy, transmit power level, signal constellation, modulation scheme, channel coding, as well as cell coverage, network topology, user preferences, communication protocols, and sensing policies, have not been cognized under a unified framework.
It is then necessary to expand the parameter space from a single ``spectrum hole'' to ``multiple parameters'' space in order to make the network more ``intelligent''.

Meanwhile, the multi-parameter space also shares the highly structured and mutually coupled  characteristics, which could be further utilized to enhance the cognition performance and reduce cognition complexity.
If sufficient  prior knowledge are available at each node, the cognition of multiple parameters will be  similar to the conventional estimation and detection problem for single parameter cognition  but with an enlarged dimension of parameter space. In this case,  modified signal processing techniques, e.g., multiple hypotheses testing, could be applied \cite{Jiachen}.
Nonetheless, in complicated networks, the prior knowledge is normally unavailable and conventional signal processing techniques can be hardly executed.
A solution for multiple parameters cognition is then to introduce the machine learning theory that is able to establish the prior knowledge base automatically from the proliferation of traffic data. The amount of data could be accumulated shortly considering the signal transmission speed over Mega or Giga bits.
In fact, machine learning has already emerged as a booming research area these years, whose core idea is to mine the ``patterns'' behind massive data and further identify or predict unknown ``patterns''.
The inherent coupling and inter-relationship of network parameters  can be recognized as the ``patterns'' through learning techniques in this context.

Practically, the parameters that determine the control and optimization policies in networks change frequently, while a small change in environmental parameters may result in a big change in the systematic behaviors \cite{SON_Swarm}. The parameter cognition should then be as adaptive and predictive as possible to cope with the dynamic characteristics of parameters.
If the cognition is able to provide accurate prediction of the network status from the patterns behind  multi-parameter space, then better quality of service can be achieved via resource pre-distribution.

In this article, we propose a new cognitive architecture based on either the conventional signal processing or the machine learning techniques, targeting at multiple parameters in future wireless communication networks.
The remainder of the article is organized as follows. We first describe the core idea and the framework of the proposed cognitive architecture.
We also provide three concrete examples that demonstrate the validity and efficacy of the architecture.
Summary and some prospects are highlighted at the end of the article.

\section{Architecture of Enhanced Multi-Parameter Cognitive Techniques}
The architecture of the proposed cognitive technique is elaborated as three main stages: parameter structuralization, multi-parameter cognition, and parameter prediction.
\subsection{Parameter Structuralization}
\begin{figure*}[ht]
  \centering
    \includegraphics[width=150mm]{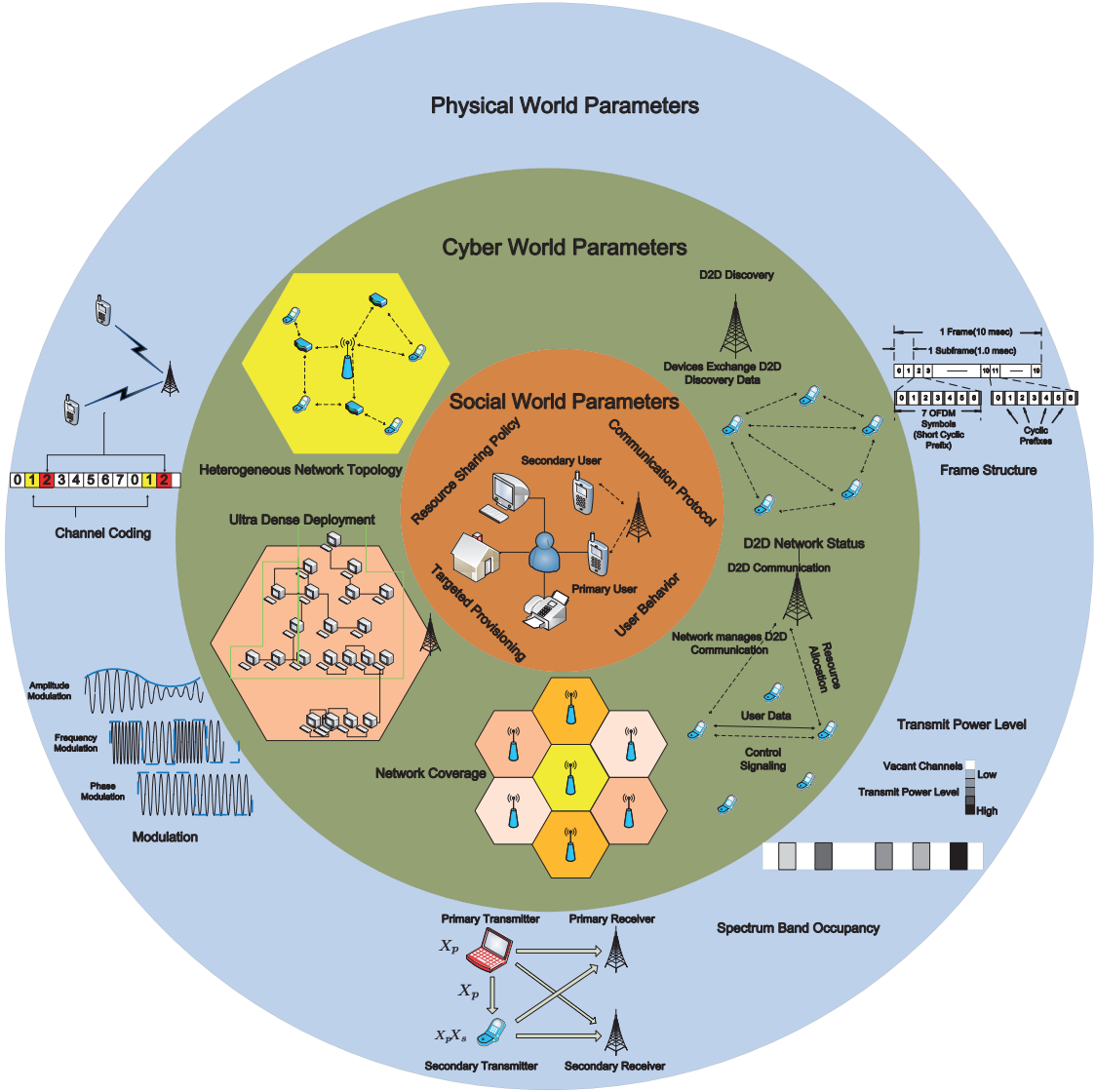}
  \caption{Enormous number of parameters involved in cognitive  communication networks.}
  \label{structure}
\end{figure*}
As illustrated in Fig. \ref{structure},  a communication network is described by gigantic amount of parameters in different layers, and cognition for as many as possible parameters at each node could   greatly release the expense of the feedback and enhance network efficiency. 

Quite many efforts have been made in parameter cognition such as channel estimation and spectrum detection, where some fine results have been established in the past decades.
However, these efforts each concerns the extraction of only an individual  parameter, ignoring the potential  relationship among different parameters, i.e., parameter structure. The structure of parameters in the same layer can be achieved through direct signal processing techniques while structure in different layers can be achieved through pattern recognition and learning techniques.  For example, the cognized symbol constellation will directly exhibit the spectrum occupancy status, and could also indicate the transmission behaviours or preferences of  users through certain data mining approach.

There are three main advantages of parameter structuralization: (i) Structuralization could improve the cognition accuracy since various parameters complement one another in the light of the ``structure''; (ii) Complexity of cognition could be reduced since smaller number of parameters are necessary to be cognized due to the correlation among them; (iii) Range of parameter cognition could be expanded because some non-cognitable parameters could be achieved by reasoning from the cognitable parameters via machine learning techniques.

\subsection{Multi-Parameter Cognition}
\begin{figure*}[ht]
  \centering
    \includegraphics[width=150mm]{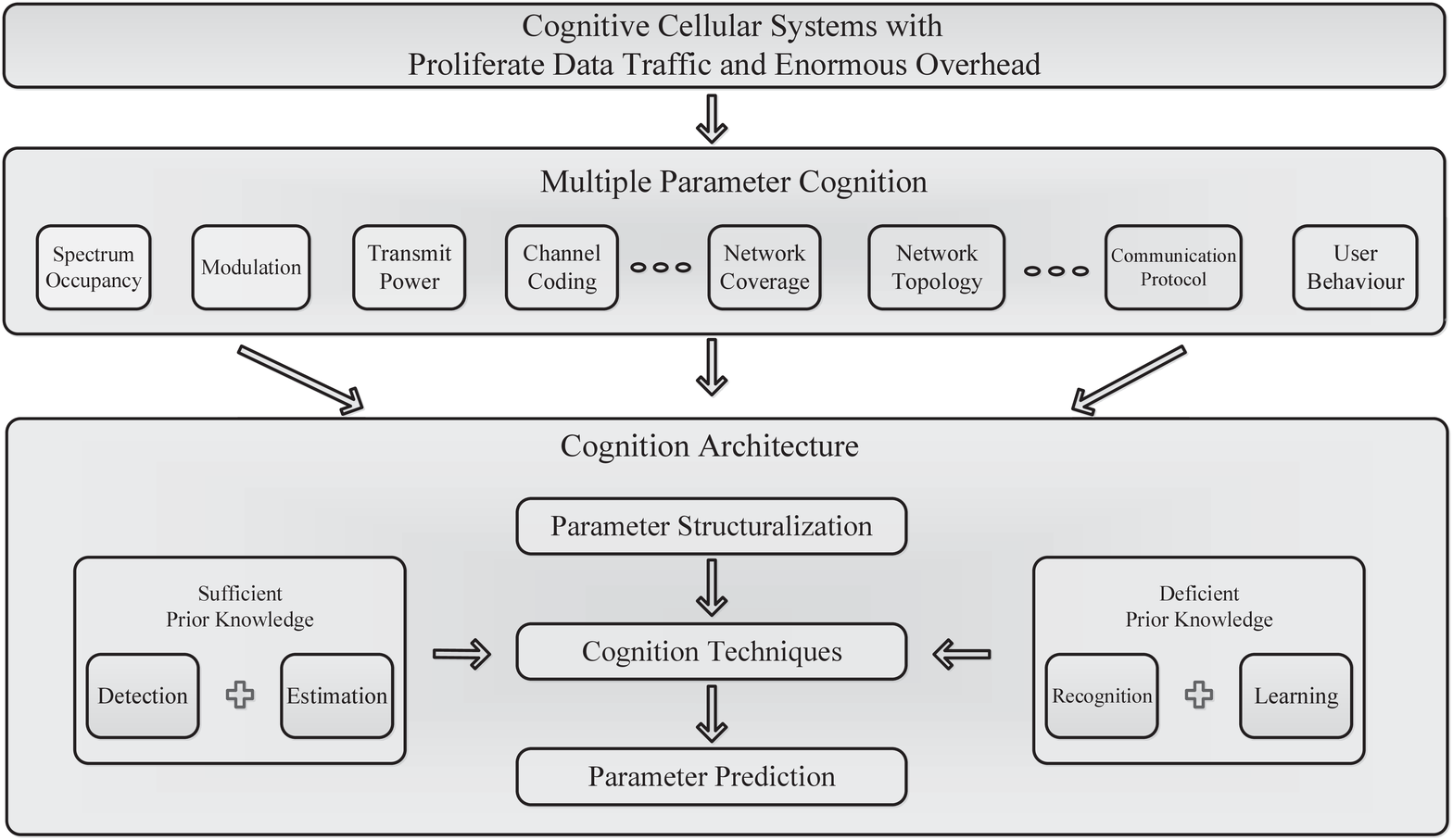}
  \caption{Framework of the multi-parameter space cognitive architecture.}
  \label{framework}
\end{figure*}
The proposed multi-parameter cognitive architecture is shown in Fig. \ref{framework}, where the most popular parameters include spectrum occupancy, transmit power, modulation, and user behaviour.

In general, parameter cognition can be categorized into two cases: (i) prior-sufficient case; (ii) prior-deficient case, depending on how much the prior knowledge is available before cognition. For the prior-sufficient case, where  the key  information, e.g., interference level, noise characteristics, channel statistics  are known to users, the multiple parameters cognition problem could be solved by conventional signal processing techniques. In this case, closed-form solutions or numerical solutions normally exist, which sometimes serve as the performance bound for other low-complex sub-optimal algorithms.

When the prior knowledge is insufficient to execute conventional signal processing algorithms, one may resort to  learning techniques, e.g., data-driven algorithms from machine learning and pattern recognition, to intelligently establish the cognition knowledge base. Indeed, some machine learning algorithms have been proposed for various  tasks in CR parameter cognition. For example,
in \cite{Huang}, a spectrum sensing engine based on support vector machine (SVM) was designed, advancing the sensing performance with smaller samples compared to the energy detector. In cooperative spectrum sensing paradigm, \cite{Lo} presents reinforce learning methods to reduce the sensing overhead by releasing network flow congestion.
 Most of these learning techniques, however, are mainly designed to obtain a single parameter, i.e., spectrum occupancy, and do not consider the learning of other system parameters.

Learning-based cognition techniques seem to be more applicable in real CR setups and  are subject to little performance downgrade in the above cases.
Nonetheless, there are still some deficiencies that are worth further attention for there is no ``free lunch''.
First, it is challenging to collect sufficient ``clean'' data for learning and recognition.
Extra overhead and high complexity for establishing and maintaining the database during the whole cognition are unavoidable.
Second, in some centralized control and optimization scenarios, learning techniques may not be adequate to cope with large-scale traffic data sets.
Hence, it may be necessary to incorporate  non-parametric learning algorithms with low computational complexity in the cognition framework.
Moreover, data-driven algorithms can be prone to data falsification and attack from malicious users, and more efforts are necessary for safety analysis and protection operations.

\subsection{Parameter Prediction}

Following  the multiple parameters cognition,  users are able to utilize internal structure as well as  underlying pattern  of parameters to  further understand the parameter evolution.
For example, channel occupancy status can be predicted by learning the traffic characteristics of licensed systems using the neural network model \cite{Tumuluru}.
The reliable prediction of channel status considerably reduces the overhead consumed in spectrum sensing for the reason that only those channels  predicted to be idle need to be sensed  in the next time-slot.
Moreover, if the unlicensed user can predict other parameters of licensed users, e.g., power level, modulation, and coding-scheme, it could then adjust the corresponding parameters such that the interference caused to the licensed user can be optimally reduced.
In intelligent communication networks, e.g., SON, predicability normally serves as the counterpart of scalability but is difficult to model due to the complexity of systems \cite{SON_Swarm}.
In this case, learning-based cognition techniques is much more applicable for parameter prediction because of its extrapolative nature and simple implementation.



\section{Examples}
In the following, we provide three representative examples of the latest research progress to show how the  signal processing and machine learning techniques could be used to cognize multiple parameters.

\subsection{Example 1: Cognition for  Transmit Power Levels}
Let us first  consider a practical CR scenario, where the licensed user could work under more than one discrete power levels, as opposed to single power level in conventional CR.
The  parameter that could be cognized is then not only the on/off status of the licensed user, but also its transmit power levels, where non-zero power levels indicates the ``on'' of licensed
user and vice versa.

\begin{figure}[t]
  \centering
    \includegraphics[width=130mm]{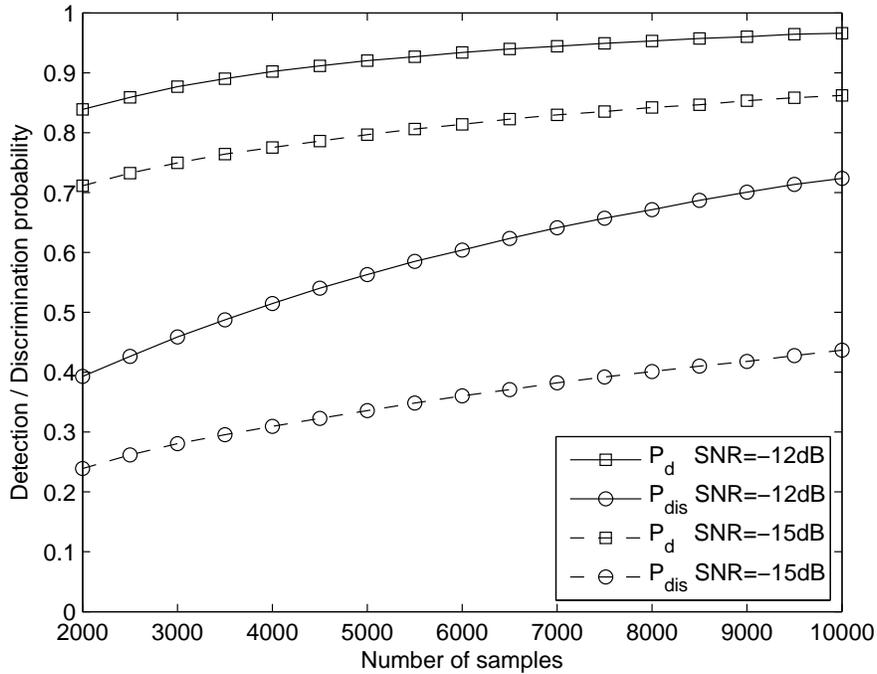}
  \caption{The theoretical performance metric of cognition versus number of samples with system parameters known in prior.}
  \label{MPTP_Performance}
\end{figure}

If the candidate power levels and noise statistics are known in prior, the spectrum occupancy as well as the current transmit power levels can be obtained via multiple hypotheses detection \cite{Jiachen}. Besides the false alarm probability and detection probability that are used to evaluate the conventional spectrum sensing, one may further define a new performance metric called discrimination probability, which describes the capability of correctly recognizing each power pattern. A numerical result is shown  in Fig. \ref{MPTP_Performance}, where the theoretical curves  can serve as an upper bound of performance for any sub-optimal detectors.

\begin{figure}[t]
  \centering
    \includegraphics[width=120mm]{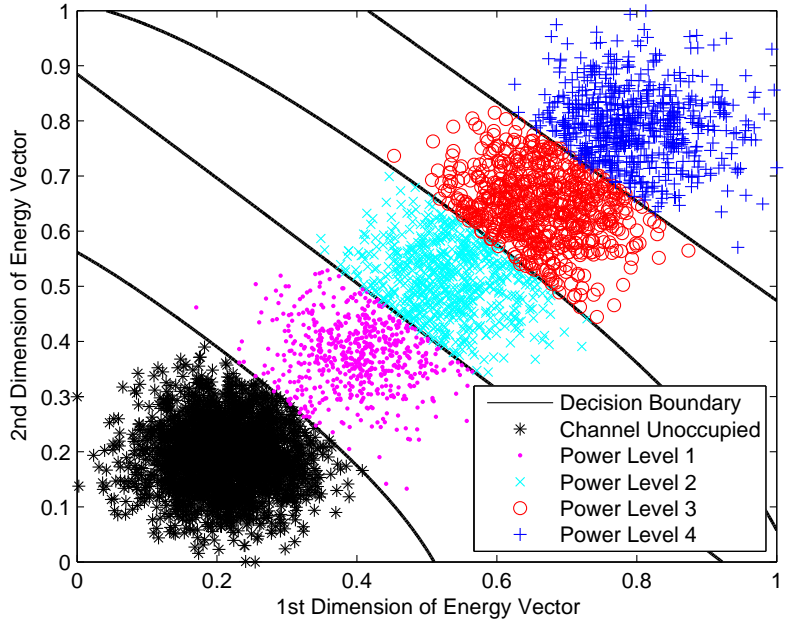}
  \caption{Clustering results and SVM trained decision boundary on normalized energy feature vectors when average signal noise ratio (SNR) is -12 dB.}
  \label{Cluster}
\end{figure}

However, the assumption that the unlicensed user fully knows all prior information cannot be true in realistic setup. In this case, the machine learning based cognition techniques, such as clustering analysis and classifier construction, could be exploited to replace the multiple hypotheses testing approach.
The unlicensed users can collect energy statistics and formulate energy feature vector through multi-slot sensing scheme \cite{Guo}, and licensed transmission patterns can be discovered by clustering analysis from a sufficient number of energy feature vectors.
Specifically, feature vectors that share the same transmit power of PU can be  grouped together so that the number of power states and the corresponding vector clusters are determined.
Hence, the transmit power levels can be evaluated by the average of the vectors in each cluster, from which the knowledge about transmit patterns and preferences, e.g., the power states of the licensed user,  the average power level of each  state, etc., are learned.
Moreover, each energy feature vector would be labeled by the cluster number it is partitioned to.
Classifiers and decision boundaries between different power states can then be constructed through supervised learning on the basis of these labeled data.
A numerical example where SVM is used to train the classifiers and perform sensing tasks is  shown in Fig. \ref{Cluster}.

Once the power level is detected, the unlicensed users are able to adjust their own transmit power such that the interference temperature for that particular power level is fulfilled,\footnote{FCC has  regulated different protection for different powered services in their recent reports \cite{FCC}.} by which means the unlicensed users could fully squeeze the tolerance of the licensed user and maximize their own throughput.

\subsection{Example 2:  Cognition for Modulation Pattern}
In this example, we address the problem of modulation recognition (MR) in CR scenario.
For conventional MR, there are  two main genres of methods: the theoretical maximum-likelihood-based methods and the statistical pattern-recognition-based methods \cite{Swami,Octavia}.

However, MR in CR differs from the convention in several ways. Firstly,  the existing MR methods assume a prior knowledge of the candidate signal constellations, which can be denoted as the modulation dictionary. It is obligatory for the dictionary to contain as many potential modulation types as possible while a redundant dictionary would definitely degrade the recognition performance, especially in low signal-to-noise ratio region. Hence the modulation dictionary should be pruned tactfully before recognition.
Secondly, the existing methods always assume the transmitter to be ``on", which is not the case in CR scenario where the channel occupancy status is also unknown and its detection is coupled with MR.
Thirdly, the existing methods do not consider different transmit power levels, which can, in fact, be coupled with modulation types to characterize the transmission behavior of the licensed users.

\begin{figure}[t]
  \centering
    \includegraphics[width=130mm]{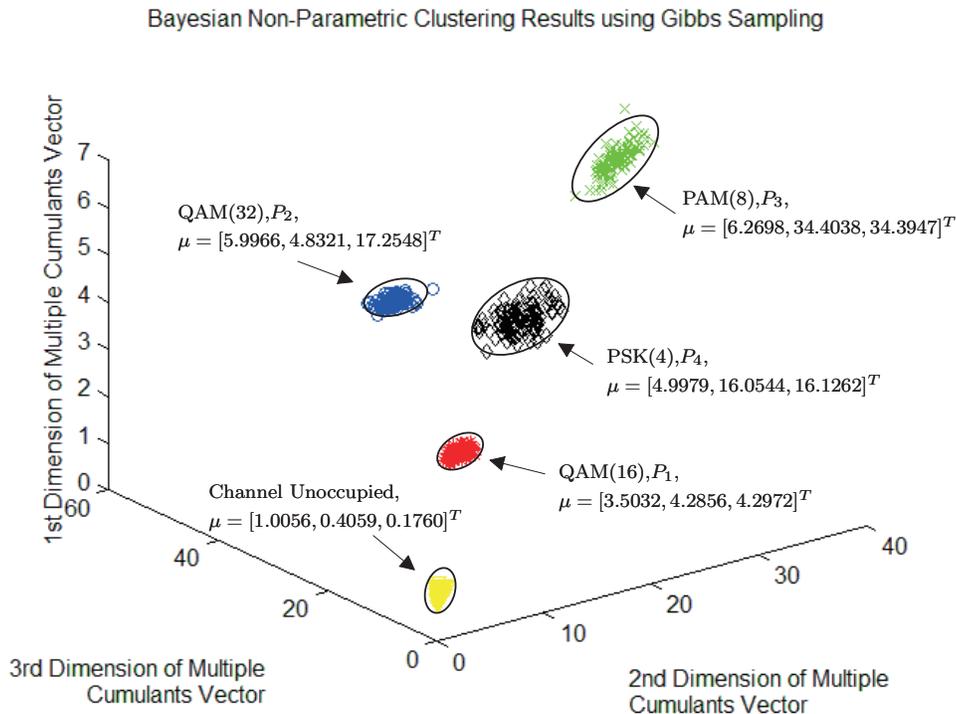}
  \caption{Pattern discovery and clustering results of multiple cumulants vectors before normlization using Dirichlet Process Gaussian Mixture Model.$\mu$ denotes the mean vector of each Gaussian component and $P_{i}$ denotes the transmit power level where $P_{1}^{2}:P_{2}^{2}:P_{3}^{2}:P_{4}^{2}=2.5:5:5.3:4$ when overall average SNR is 10 $dB$ and the number of samples is 100.}
  \label{DPGMM}
\end{figure}

We provide one numerical example in Fig. \ref{DPGMM} to shed lights on how to obtain  the coupled parameters from unsupervised learning approach.
Higher order statistics (HOS) are used as the feature for recognition because they characterize the  distribution  shape of  noisy baseband samples with low complexity \cite{Swami}.
We introduce a new concept ``Modulation Pattern" to denote the combination of the modulation type and the transmit power level.
We then organize the estimates of  cumulants of different orders and lags as a multiple cumulant  vector, serving as the input feature of the machine learning algorithms. It can be proved that the feature vector is  a multivariate asymptotic Gaussian approximation of cumulants \cite{Dandawate}.
Specially, the first cumulant in the vector should be the second-order cumulant in order to indicate the energy information.
In order to construct the modulation dictionary and identify different ``modulation patterns'', we exploit Dirichlet Process Gaussian Mixture Model (DPGMM), a type of unsupervised clustering analysis,  to construct dynamic and flexible statistical representations for the training data. We formulate a mixture model in which the number of mixture components is infinite and is not required at the beginning of clustering.
After the convergence of DPGMM \cite{Rasmussen}, the cumulant vectors that aggregate around the origin of coordinate system  represent  the ``noise constellation".
Detection of cumulant vectors belonging to ``noise constellation'' reveals that only noise signals are received and no symbols are transmitted.
The variance of noise can be evaluated from the second-order cumulant of ``noise constellation''.
Furthermore, the average vector of any other cluster is evaluated.
The average vectors are used to identify the alive modulation types and establish the minimal dictionary following the correspondence of HOS and modulations.
Consequently, new observations of cumulant vector can be efficiently classified and predicted using the predictive probability distribution constructed by DPGMM.
The dictionary can also be updated along with the update of posterior predictive distributions adaptively.

The simulation results in Fig. \ref{DPGMM} show the efficacy of the modulation recognition and indicate superior discrimination capability compared with the pure pattern recognition approach \cite{Swami}.
Moreover, Fig. \ref{DPGMM} also demonstrates that four modulation types as well as the channel idle status can be specified automatically, along which the transmit power level can be cognized as well.
Furthermore, the modulation preference of certain user is understood from each component of the predictive distribution.

\subsection{Example 3: Prediction for Spectrum Environment}
In the last example, we evaluate the prediction of  multi-channels spectrum occupancy using machine learning techniques.
Due to energy and hardware constraints, unlicensed users may not be able to perform spectrum sensing at all channels.
This can be relieved by predicting the channel occupancies before each sensing time slot starts.
As modeled in \cite{Zhihui}, the state of different channels and the time the licensed users spend dwelling on each state are assumed to be independent.
Moreover, the change of spectrum environment can be considered as the process of license channel vacancy and occupancy states appear alternately. 
We introduce throughput as the evaluation criterion of the proposed learning-based cognitive strategy.

\begin{figure}[ht]
  \centering
    \includegraphics[width=130mm]{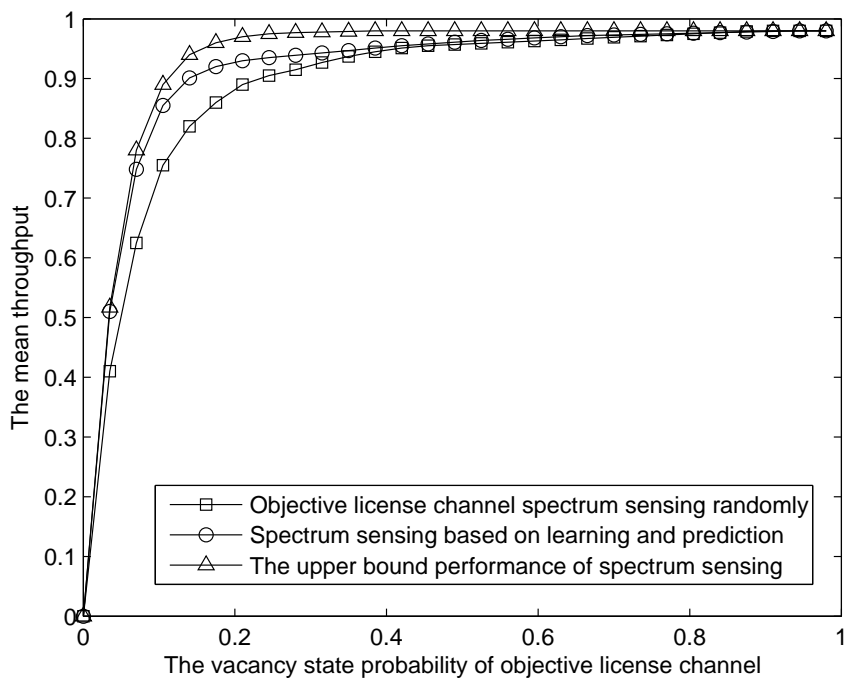}
  \caption{The performance of mean throughput versus the probability of vacancy state when the number of channels is $25$ and the normalized channel capacity is $1$bit/s.}
  \label{Prediction}
\end{figure}

The multi-channels spectrum sensing based on learning and prediction can be performed via three steps.
Primarily, the parameters of the learning model are estimated by the historical information embedded in the previous observations database.
Subsequently, given a new observation of the channel occupancy, the next time slot can be predicted using the learning model. 
In the meanwhile the probability of vacancy status of each channel can be ordered from high to low.
Lastly the unlicensed user is able to efficiently perform spectrum sensing  in accordance with the order and update the channel status database with the true channel status detected.
The average throughput can be derived in terms of the mean error prediction probability and the probability of vacancy state for objective channels in closed-form \cite{Zhihui}.
Simulation results in Fig. \ref{Prediction} demonstrate that the multi-channels spectrum sensing and prediction based on machine learning techniques achieve better performance in terms of throughput, compared with random selection of license channels.

Besides the enhancement of spectrum sensing efficiency, there are a few other benefits gained from learning and prediction of multiple channels occupancy.
Firstly, prediction results can also be utilized to foresee the traffic flow of cellular networks consist of certain users at certain spectrum sub-bands. Hence the network congestion may be detected in advance.
Moreover, prediction of channel occupancies makes it possible for predicting  transmission behaviour of users.

\section{Summary and Prospects}
In this article, we introduce a concrete  multi-parameter cognitive architecture for future wireless communication systems that contains three key stages.
Our efforts  prompt a compromising but necessary way towards the ultimate goal of CR, i.e., ``full cognition'', which still stays in the conceptual level.
Examples demonstrate that the cognition in  multiple parameters space can be rather different from conventional single parameter cognition and  reveal fundamental insights in the proposed cognitive architecture.
The future research could include how to utilize the cognized parameters to preserve quality of service of the network in terms of data traffic, latency, overhead, etc.

\linespread{1.5}

\end{document}